\documentclass{article}
\usepackage{spconf,amsmath,graphicx}
\usepackage{algorithm} 
\usepackage{algpseudocode} 
\usepackage[table,xcdraw]{xcolor}
\usepackage{comment}

\title{ mixPGD: Hybrid Adversarial Training for Speech Recognition Systems}
\begin{document}
\name{Aminul Huq \qquad Weiyi Zhang \qquad Xiaolin Hu$^{\star}$}
  
  \address{Department of Computer Science and Technology, \\
  State Key Laboratory of Intelligent Technology and Systems,\\ Tsinghua Laboratory of Brain and Intelligence (THBI),\\ IDG/McGovern Institute of Brain Research,\\ Tsinghua University, Beijing 10084,China \\
  huqa10@mail.tsinghua.org.cn, wy-zhang19@mails.tsinghua.edu.cn, xlhu@tsinghua.edu.cn } 

\maketitle
\begin{abstract}
Automatic speech recognition (ASR) systems based on deep neural networks are weak against adversarial perturbations. We propose mixPGD adversarial training method to improve the robustness of the model for ASR systems. In standard adversarial training, adversarial samples are generated by leveraging supervised or unsupervised methods. We merge the capabilities of both supervised and unsupervised approaches in our method to generate new adversarial samples which aid in improving model robustness. Extensive experiments and comparison across various state-of-the-art defense methods and adversarial attacks have been performed to show that mixPGD gains 4.1\% WER of better performance than previous best performing models under white-box adversarial attack setting. We tested our proposed defense method against both white-box and transfer based black-box attack settings to ensure that our defense strategy is robust against various types of attacks. Empirical results on several adversarial attacks validate the effectiveness of our proposed approach.
\end{abstract}
\section{Introduction}
\label{sec:intro}

Deep neural networks (DNNs) from its very inception have been able to prove itself a very powerful tool. Its capability to find out important features has made it popular to use in every known scenario. We can see the presence of DNN in the field of classification tasks \cite{druzhkov2016survey}, segmentation \cite{yin2021end}, medical analysis \cite{litjens2017survey}, self-driving vehicles \cite{ni2020survey}, automatic speech recognition systems \cite{kumar2018survey}, sentiment analysis \cite{zhang2018deep}, and many more. Many real-life applications based on DNN have been deployed as well. 

\let\thefootnote\relax\footnotetext{$\star$ Corresponding author.}

One of the most used deep learning application in the current world is automatic speech recognition (ASR) systems. Speech is the primary way of communicating across all animals on our planet. That
is why making intelligent devices understand human speech is a popular and long-cherished task.
ASR systems take input as raw audio waveform and produce transcriptions of the same. Previously using the Hidden Markov Model was the most used method but recently various deep learning approaches have become
widely popular and more accepted. Models like  Listen Attention and Spell \cite{ChanJLV16}, Wav2Letter \cite{SchneiderBCA19} etc. are quite popular for ASR tasks. Nowadays to automate our daily life we have been using various voice assistant and voice controlled devices.

In spite of the success of deep learning methods, it has been proven recently that these models are unreliable. Recent studies have shown the existence of adversarial perturbations which are capable of harming the performance of these DNNs by adding imperceptible noise to the input data. For audio data, perturbations can not be heard over normal speech. To protect DNNs from any adversarial attacks, study of these adversarial attacks and defense techniques are necessary. 
\begin{figure}[t]
  \centering
  \includegraphics[width=1\linewidth]{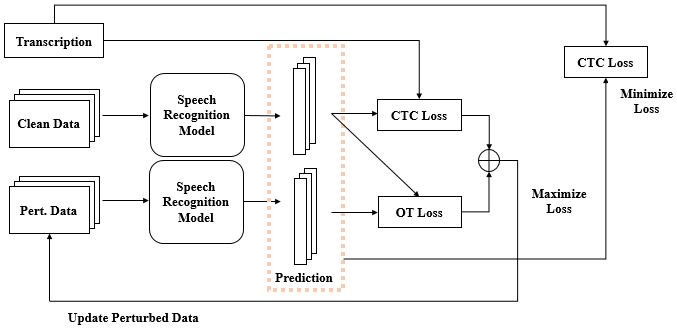}
  \caption{Schematic diagram of mixPGD adversarial training.}
\end{figure}

One of the most effective defense strategy to mitigate the effect of adversarial perturbations is adversarial training. There are also other approaches like detection based methods proposed by \cite{YangQCML20, Ma,JayashankarRM20} and preprocessing based methods like \cite{YangLCS19,AndronicKRRS20}. These methods have limitations like the detection based methods discards the data during inference time and the preprocessing based methods are not always capable of cleaning the perturbations completely. During the adversarial training approach both original clean and adversarial samples are used to train the model. In the field of ASR systems few adversarial training based defense strategies have been explored. S. Sun et al used Fast Gradient Sign Method (FGSM) based adversarial training to train a model \cite{SunYOHX18}. However, they did not mention their approaches performance under diverse adversarial attacks and attack scenarios like black-box attacks. Most research work focused on either detection based methods or on preprocessing methods. 

The motivation behind our approach is based on the fact that most of the existing work focuses on either supervised approach like FGSM and PGD adversarial training or unsupervised approach like Feature Scattering. We propose a new type of adversarial sample generation technique namely mixPGD approach. We propose to craft adversarial samples by merging both supervised and unsupervised loss techniques which would be used to perform adversarial training. Our new adversarial sample generation technique is based on incorporating both the supervised and unsupervised loss calculating schemes to generate new hybrid adversarial samples. Figure 1 shows a schematic diagram of our method. We explore our methods effectiveness under a set of diverse adversarial attacks in both white-box and black-box attack settings. We also explore which unsupervised loss estimation method would aid in this method. 

\section{MixPGD Adversarial Training}

Most of the adversarial attacks like FGSM, PGD etc are supervised ones. These approaches have the benefit of generating stronger adversarial samples. However these approaches do not consider inter-sample relationship and is susceptible to label-leaking. In order to solve these problems unsupervised adversarial sample generation methods were proposed like Feature Scattering (FS) \cite{ZhangW19}. This method takes time and not always generates stronger adversarial samples. We wanted to create a method which embodied the capabilities of both supervised and unsupervised methods. Intuitively, if both of these methods are merged together we would get stronger adversarial samples which in turn will aid us in performing adversarial training.

Adversarial samples are generated by using the loss value with respect to the input data for getting the gradient information. In supervised scenarios, cross-entropy (for classification), connectionist temporal classfication (for speech recognition model) etc. are used while in unsupervised approach no labels are used but rather the difference between clean samples prediction and adversarial samples prediction are used. We can define connectionist temporal classfication (CTC) \cite{GravesFGS06} loss between the original label and the prediction of the model as follows,
\begin{equation}
    L_{\text{CTC}} (f(x),y)
\end{equation}
where $x$ is the audio input which is fed to the speech recognition model $f$ and $y$ is the corresponding transcription. This equation is a supervised loss function as we are using the original label of the data. Generally in supervised adversarial sample generation techniques like FGSM and PGD we try to maximize Equation 1. For the unsupervised loss we are using the concept of optimal transport (OT) theory. This method was used as an unsupervised loss function in \cite{ZhangW19}. The authors in this paper tried to maximize the loss between the predicted label  and the adversarial label to generate stronger adversarial samples. In this case the adversarial sample is initialized with random noise. The OT distance can be represented as,
\begin{equation}
    L_{\text{OT}} = \underset{T}{\text{min}} \hspace{3.5pt} (T \cdot C)
\end{equation}
where $T$ is a matrix which helps to solve the OT problem and is calculated in this paper using Sinkhorn's approach\cite{Cuturi13}. $C$ is the transportation cost matrix. We used the cosine distance between the predictions of the clean sample $f(x)$ and the prediction of the adversarial sample $f(\tilde{x}^t)$ It is defined as:

\begin{equation}
       C = 1 - \frac{f(x)^T f(\tilde{x}^t)}{||f(x)||_2 ||f(\tilde{x}^t)||_2} .
\end{equation}

\begin{algorithm}[t]
	\caption{mixPGD Adversarial Training} 
	\textbf{Inputs:} Training data $\{x_i,y_i\}_{i=1,..,n}$, outer iteration number $T_o$, inner iteration number $T_i$, maximum perturbation $\epsilon$, step size $\eta_1$ and $\eta_2$, network architecture parametrized by $\theta$.\\
    \textbf{Outputs:} Robust Speech Recognition Model, $f_\theta$
	\begin{algorithmic}[1]
		\For {$t=1,2,\ldots T_o$}
		\State Uniformly sample a batch of training data $B^{(t)}$
			\For { $x_i \in B^{(t)}$}
			\State $x_i^{'} \longleftarrow x_i + 0.0001 \cdot N(0,I)$, where $N(0,I)$ is the Gaussian distribution with zero mean and identity variance.
			    \For{$s = 1 ,\dots , T_i$}
			    \State generate adv sample using Eqn. (5)
			    
%
			    
			    \EndFor
		   \EndFor
		   \State $\theta \longleftarrow \theta - \eta_2 \sum_{x_i \in B^{(t)}}  \nabla_\theta ( 
    L(f_{\theta}(x_i^{'}),y_i)  ) $
    \EndFor
	\end{algorithmic} 
\end{algorithm}
We propose to combine both of the supervised loss and unsupervised loss, i.e, Equation 1 and Equation 2 and use it to craft new adversarial samples. By using these new samples we plan to perform adversarial training to improve the robustness of our speech recognition model. Formally this loss function can be described as:

\begin{equation}
    L_{\text{new}} = L_{\text{CTC}} + \beta L_{{\text{OT}}}.
\end{equation}
Here, $\beta$ is a weighting factor balancing the supervised and unsupervised losses. In our experiments we found out that setting $\beta=1$ was good enough. By using Equation 4 we can generate hybrid adversarial samples. This is an iterative process and it can be described as, 

\begin{equation}
    \Tilde{x}^{(t+1)} = \tilde{x}^t + \epsilon \cdot \text{sign} ( \nabla_{x} (L_{\text{new}})).
\end{equation}
Here, $\tilde{x}$ represents the mixPGD adversarial example of original input $x$ and $y$ is the corresponding transcription. The proposed approach is an iterative one and $t$ refers to the iteration number. The adversarial samples of the previous iteration are used to generate samples in the next iteration. The task of the adversarial training is to minimize the CTC loss between the adversarial sample and target. The adversarial sample is generated by taking the gradients of the summed loss. The overall approach is describe in Algorithm 1.

\section{Experiment}

\subsection{Speech Recognition Model}
\begin{figure}[t]
  \centering
    {\includegraphics[width=0.95\linewidth]{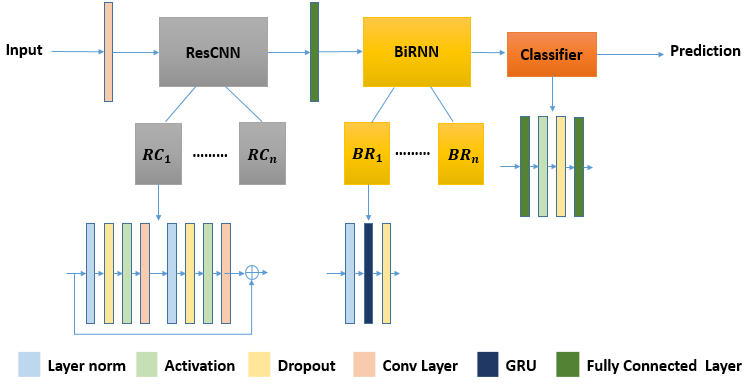}}
  \caption{Speech recognition model.}
\end{figure}
At the very beginning of our speech recognition model a single convolution layer exists that takes in the input data. This speech recognition model has two main deep neural network components. One is Residual Convolutional Neural Networks (ResCNN) and the other part is Bi-directional Recurrent Neural Networks (Bi-RNN). We used the Gated Recurrent Unit (GRU) in this experiment. In the end, there are some fully connected layers so that the input can be classified into character per time step. To decode the output to text, we used a greedy decoder in this experiment. There are various other decoders that may perform better but the main purpose of this experiment is to evaluate the robustness of the model that is why this decoder has been used for its simplicity. Figure 2  provides a detailed view of the speech recognition model$^1$. This model achieved 28.75\% of WER. 

\footnote{$^1$ https://www.assemblyai.com/blog/end-to-end-speech-recognition-pytorch/}

\subsection{Experimental Setup}
To perform automatic speech recognition we used the popular speech recognition dataset named Librispeech\cite{PanayotovCPK15}. There are more than 100 hours of speech data recorded here at 16KHz frequency. All the training and testing data are in raw audio format. As raw audio is very inefficient to use to train a speech recognition model, we take the mel-spectrogram features of the audio data for both training and testing. The speech recognition model is trained using AdamW \cite{LoshchilovH19} optimizer and One Cycle Learning Rate \cite{smith2019super} scheduler. The initial learning rate was set at 0.0005 with batch size of 10. The speech recognition model is trained on 25 epochs. We used two different evaluation metrics here, one is character error rate (CER) and the other is word error rate (WER). In this experiment, we trained five different speech recognition models including ours to compare which model provides better robustness. The speech recognition model which is trained without any defense strategies is called a standard trained model. Other speech recognition models all have some defense strategies included in them like FGSM, PGD adversarial training, Feature Scattering and our proposed method. We tested these models against FGSM, MIFGSM, and PGD adversarial attacks. During training and testing the maximum value for allowed perturbation $\epsilon$ was 0.00004 with step sizes of $\epsilon/4$. In \cite{JatiHPPAN21}  the authors used a small perturbation size similar to ours. We created the same number of adversarial samples as the original data for both training and testing purposes. We used four NVIDIA GeForce GTX 1080 Ti GPUs for training and testing purpose. 

\begin{table*}[h]
\centering
\caption{CER(\%) and WER(\%) comparison of different methods under white-box attacks.}
\label{tab:my-table}
\resizebox{\textwidth}{!}{%
\begin{tabular}{|c|c|c|c|c|c|c|c|c|c|c|}
\hline
 &
  \multicolumn{2}{c|}{\begin{tabular}[c]{@{}c@{}}Standard \\ Training\end{tabular}} &
  \multicolumn{2}{c|}{\begin{tabular}[c]{@{}c@{}}FGSM\\ adv\end{tabular}} &
  \multicolumn{2}{c|}{\begin{tabular}[c]{@{}c@{}}PGD\\ adv\end{tabular}} &
  \multicolumn{2}{c|}{\begin{tabular}[c]{@{}c@{}}Feature \\ Scattering\end{tabular}} &
  \multicolumn{2}{c|}{\begin{tabular}[c]{@{}c@{}}mixPGD\\ adv\end{tabular}} \\ \hline
       & CER      & WER    & CER      & WER    & CER      & WER    & CER      & WER    & CER               & WER             \\ \hline
Clean  & 11.3370 & 28.78 & 10.8772 & 33.70 & 10.6461 & 33.16 & 9.6515 & 30.01 & \textbf{9.2731} & \textbf{29.02} \\ 
FGSM   & 15.4303 & 48.70 & 13.1881 & 40.23 & 12.7255 & 39.20 & 12.7788 & 39.08 & \textbf{11.2406} & \textbf{35.07} \\ 
MIFGSM & 17.1673 & 57.29 & 13.1366 & 39.75 & 12.8168 & 39.54 & 13.4278 & 41.02 & \textbf{11.2915} & \textbf{35.15} \\ 
PGD20  & 21.8359 & 69.69 & 13.1991 & 40.29 & 12.8599 & 39.59 & 14.2479 & 45.34 & \textbf{11.3015} & \textbf{35.29} \\ 
PGD100 & 24.2174 & 75.61 & 13.1569 & 40.45 & 12.8633 & 41.02 & 14.5591 & 47.60 & \textbf{11.4232} & \textbf{35.39} \\ \hline
\end{tabular}%
}
\end{table*}

\subsection{Performance Under White-box Attacks}
We performed a diverse set of adversarial attacks in white-box attack settings on popular adversarial defense strategies and our proposed method to compare which defense strategy is more robust. We performed FGSM, MIFGSM, PGD20 and PGD100 adversarial attack. All adversarial attacks were performed under a fixed perturbation budget $\epsilon = 0.00004$. We evaluated different models performance using character error rate (CER) and word error rate (WER). The results are shown in Table 1. Each row presents standard training, FGSM adversarial training, PGD adversarial training, Feature Scattering and mixPGD adversarial training models performance against different attacks shown in the very first column. We can see that standard trained model had been affected the most by adversarial attacks, as it had no defense mechanism. FS method achieved good results against FGSM attack but its performance was not good under PGD attack. FS approach does not generate strong adversarial samples with respect to the original labels so it may under-perform against stronger adversaries. Also another important factor is that for ASR systems we have to use a decoder to generate the transcripts this might cause reduced performance. Under all attacks PGD adversarial training performed well. However, our mixPGD based adversarial training achieved better CER and WER score against all type of attacks since this utilizes both supervised and unsupervised loss. It gained 4.1\% WER of better performance than PGD adversarial training against PGD20 adversarial attack. 

\subsection{Performance Under Transfer Based Black-box Attacks}

\begin{table}[t]
\centering
\caption{WER(\%) comparison of different methods under transfer based black-box attacks.}
\label{tab:my-table}
\begin{tabular}{|c|c|c|c|}
\hline
{\color[HTML]{000000} } &
  {\color[HTML]{000000} FGSM} &
  {\color[HTML]{000000} MIFGSM} &
  {\color[HTML]{000000} PGD50} \\ \hline
{\color[HTML]{000000} \begin{tabular}[c]{@{}c@{}}Standard\\ Training\end{tabular}} & 
  {\color[HTML]{000000} 38.93} &
  {\color[HTML]{000000} 41.44} &
  {\color[HTML]{000000} 46.43} \\ \hline
{\color[HTML]{000000} \begin{tabular}[c]{@{}c@{}}FGSM \\ adv\end{tabular}} &
  {\color[HTML]{000000} 33.93} &
  {\color[HTML]{000000} 33.95} &
  {\color[HTML]{000000} 34.05} \\ \hline
{\color[HTML]{000000} \begin{tabular}[c]{@{}c@{}}PGD\\ adv\end{tabular}} &
  {\color[HTML]{000000} 33.25} &
  {\color[HTML]{000000} 33.28} &
  {\color[HTML]{000000} 33.32} \\ \hline
{\color[HTML]{000000} \begin{tabular}[c]{@{}c@{}}Feature\\ Scattering\end{tabular}} &
  {\color[HTML]{000000} 57.50} &
  {\color[HTML]{000000} 58.16} &
  {\color[HTML]{000000} 58.71} \\ \hline
{\color[HTML]{000000} \begin{tabular}[c]{@{}c@{}}mixPGD\\ adv\end{tabular}} &
  {\color[HTML]{000000} \textbf{29.26}} &
  {\color[HTML]{000000} \textbf{29.36}} &
  {\color[HTML]{000000} \textbf{29.38}} \\ \hline
\end{tabular}
\end{table}

In transfer based black-box adversarial attack settings as any information about the target model is not known to the adversary they use a surrogate model to generate adversarial samples and use it to attack the model. To mimic the same effect we used a different speech recognition to generate FGSM, MIFGSM and PGD50 adversarial samples. We compared our models performance against standard training, FGSM, PGD adversarial training and FS method on these adversarial samples. The results can be found in Table 2. Each row represents different defense models WER performance against different adversarial attacks shown in the columns. We can see that under this scenario our proposed method out-performed other defense models as well. 

\subsection{Impact of Unsupervised Loss Calculation.}

We study the importance of choosing the appropriate unsupervised loss calculation technique in this section. We compared the Kullback–Leibler (KL) divergence the OT distance. We performed mixPGD adversarial training twice and used these approaches one at a time. Then performed different white-box adversarial attacks on them. From the results displayed in Table 3 we can see that using OT distance would be the better choice than using KL divergence. 

\begin{table}[t]
\centering
\caption{WER(\%) Comparison between unsupervised loss calculation}
\label{tab:my-table}
\begin{tabular}{|c|c|c|c|c|}
\hline
         & FGSM   & MIFGSM & PGD20  & PGD100 \\ \hline
KL Div  & 39.59 & 39.76 & 40.13 & 40.35 \\ 
OT      & \textbf{35.07} & \textbf{35.15} & \textbf{35.29} & \textbf{35.39} \\ \hline
\end{tabular}
\end{table}

\section{Conclusion}
Adversarial samples based on audio data can be very troublesome in modern days. That is why it is very important to work on an effective defense strategy to protect deep learning models from adversarial attacks. In this paper, we discuss a new adversarial training based defense that generates adversarial examples in a new way. The generated adversarial samples contain the capabilities of both supervised and unsupervised approach which was not considered before by any other researchers. We perform white-box and black-box attacks on our proposed method and other popular defense strategies as well and compare the results. After extensive experimentation, we found out that our proposed approach performed much better than all other popular methods.\\
\textbf{Acknowledgements} This work was supported in part by the National Natural Science Foundation of China (Nos. U19B2034, 61836014 and 61620106010).

\bibliographystyle{IEEEbib}
\bibliography{strings,refs}

\begin{thebibliography}{10}

\bibitem{druzhkov2016survey}
Pavel~Nikolaevich Druzhkov and Valentina~Dmitrievna Kustikova,
\newblock ``A survey of deep learning methods and software tools for image
  classification and object detection,''
\newblock {\em Pattern Recognition and Image Analysis}, vol. 26, no. 1, pp.
  9--15, 2016.

\bibitem{yin2021end}
Zi~Yin, Valentin Yiu, Xiaolin Hu, and Liang Tang,
\newblock ``End-to-end face parsing via interlinked convolutional neural
  networks,''
\newblock {\em Cognitive Neurodynamics}, vol. 15, pp. 169--179, 2021.

\bibitem{litjens2017survey}
Geert Litjens, Thijs Kooi, Babak~Ehteshami Bejnordi, Arnaud Arindra~Adiyoso
  Setio, Francesco Ciompi, Mohsen Ghafoorian, Jeroen Awm Van~Der Laak, Bram~Van
  Ginneken, and Clara~I S{\'a}nchez,
\newblock ``A survey on deep learning in medical image analysis,''
\newblock {\em Medical Image Analysis}, vol. 42, pp. 60--88, 2017.

\bibitem{ni2020survey}
Jianjun Ni, Yinan Chen, Yan Chen, Jinxiu Zhu, Deena Ali, and Weidong Cao,
\newblock ``A survey on theories and applications for self-driving cars based
  on deep learning methods,''
\newblock {\em Applied Sciences}, vol. 10, no. 8, pp. 2749, 2020.

\bibitem{kumar2018survey}
Akshi Kumar, Sukriti Verma, and Himanshu Mangla,
\newblock ``A survey of deep learning techniques in speech recognition,''
\newblock in {\em International Conference on Advances in Computing,
  Communication Control and Networking (ICACCCN)}. IEEE, 2018, pp. 179--185.

\bibitem{zhang2018deep}
Lei Zhang, Shuai Wang, and Bing Liu,
\newblock ``Deep learning for sentiment analysis: A survey,''
\newblock {\em Wiley Interdisciplinary Reviews: Data Mining and Knowledge
  Discovery}, vol. 8, no. 4, pp. e1253, 2018.

\bibitem{ChanJLV16}
William Chan, Navdeep Jaitly, Quoc~V. Le, and Oriol Vinyals,
\newblock ``Listen, attend and spell: {A} neural network for large vocabulary
  conversational speech recognition,''
\newblock in {\em {IEEE} International Conference on Acoustics, Speech and
  Signal Processing (ICASSP)}, 2016, pp. 4960--4964.

\bibitem{SchneiderBCA19}
Steffen Schneider, Alexei Baevski, Ronan Collobert, and Michael Auli,
\newblock ``wav2vec: Unsupervised pre-training for speech recognition,''
\newblock in {\em INTERSPEECH}. 2019, pp. 3465--3469, {ISCA}.

\bibitem{YangQCML20}
Chao{-}Han~Huck Yang, Jun Qi, Pin{-}Yu Chen, Xiaoli Ma, and Chin{-}Hui Lee,
\newblock ``Characterizing speech adversarial examples using self-attention
  u-net enhancement,''
\newblock in {\em {IEEE} International Conference on Acoustics, Speech and
  Signal Processing (ICASSP)}, 2020, pp. 3107--3111.

\bibitem{Ma}
Pingchuan Ma, Stavros Petridis, and Maja Pantic,
\newblock ``Detecting adversarial attacks on audio-visual speech recognition,''
\newblock {\em CoRR}, vol. abs/1912.08639, 2019.

\bibitem{JayashankarRM20}
Tejas Jayashankar, Jonathan~Le Roux, and Pierre Moulin,
\newblock ``Detecting audio attacks on {ASR} systems with dropout
  uncertainty,''
\newblock in {\em Interspeech 2020}, 2020, pp. 4671--4675.

\bibitem{YangLCS19}
Zhuolin Yang, Bo~Li, Pin{-}Yu Chen, and Dawn Song,
\newblock ``Characterizing audio adversarial examples using temporal
  dependency,''
\newblock in {\em 7th International Conference on Learning Representations
  (ICLR)}, 2019.

\bibitem{AndronicKRRS20}
Iustina Andronic, Ludwig K{\"{u}}rzinger, Edgar Ricardo~Chavez Rosas, Gerhard
  Rigoll, and Bernhard~U. Seeber,
\newblock ``{MP3} compression to diminish adversarial noise in end-to-end
  speech recognition,''
\newblock in {\em Speech and Computer - 22nd International Conference,
  (SPECOM)}. 2020, vol. 12335, pp. 22--34, Springer.

\bibitem{SunYOHX18}
Sining Sun, Ching{-}Feng Yeh, Mari Ostendorf, Mei{-}Yuh Hwang, and Lei Xie,
\newblock ``Training augmentation with adversarial examples for robust speech
  recognition,''
\newblock in {\em 19th Annual Conference of the International Speech
  Communication Association}. 2018, pp. 2404--2408, {ISCA}.

\bibitem{ZhangW19}
Haichao Zhang and Jianyu Wang,
\newblock ``Defense against adversarial attacks using feature scattering-based
  adversarial training,''
\newblock in {\em 32nd Annual Conference on Neural Information Processing
  Systems (NeurIPS)}, 2019, pp. 1829--1839.

\bibitem{GravesFGS06}
Alex Graves, Santiago Fern{\'{a}}ndez, Faustino~J. Gomez, and J{\"{u}}rgen
  Schmidhuber,
\newblock ``Connectionist temporal classification: labelling unsegmented
  sequence data with recurrent neural networks,''
\newblock in {\em International Conference on Machine Learning}, 2006, vol.
  148, pp. 369--376.

\bibitem{Cuturi13}
Marco Cuturi,
\newblock ``Sinkhorn distances: Lightspeed computation of optimal transport,''
\newblock in {\em 27th Annual Conference on Neural Information Processing
  Systems (NeurIPS)}, 2013, pp. 2292--2300.

\bibitem{PanayotovCPK15}
Vassil Panayotov, Guoguo Chen, Daniel Povey, and Sanjeev Khudanpur,
\newblock ``Librispeech: An {ASR} corpus based on public domain audio books,''
\newblock in {\em {IEEE} International Conference on Acoustics, Speech and
  Signal Processing (ICASSP)}. 2015, pp. 5206--5210, {IEEE}.

\bibitem{LoshchilovH19}
Ilya Loshchilov and Frank Hutter,
\newblock ``Decoupled weight decay regularization,''
\newblock in {\em 7th International Conference on Learning Representations
  (ICLR)}, 2019.

\bibitem{smith2019super}
Leslie~N Smith and Nicholay Topin,
\newblock ``Super-convergence: Very fast training of neural networks using
  large learning rates,''
\newblock in {\em Artificial Intelligence and Machine Learning for Multi-Domain
  Operations Applications}. International Society for Optics and Photonics,
  2019, vol. 11006, p. 1100612.

\bibitem{JatiHPPAN21}
Arindam Jati, Chin{-}Cheng Hsu, Monisankha Pal, Raghuveer Peri, Wael
  AbdAlmageed, and Shrikanth Narayanan,
\newblock ``Adversarial attack and defense strategies for deep speaker
  recognition systems,''
\newblock {\em Computer Speech and Language}, vol. 68, pp. 101199, 2021.

\end{thebibliography}

\end{document}